\documentclass[psfig,graphicx,useAMS,a4paper]{iopart}
\usepackage{epsfig,txfonts,iopams,a4wide}

\begin{document}

\title[Cosmic Rays Origin and Propagation Model]{Cosmic Ray Origin and Propagation Model}

\author[A.S. Popescu]{Adrian Sabin Popescu\dag\
\footnote[3]{sabinp@aira.astro.ro}}
\address{\dag\ Astronomical Institute of Romanian Academy, Str. Cutitul de Argint 5, RO-040557 
Bucharest, Romania}

\begin{abstract}
It is presumed that the observed cosmic rays up to about $3\times 10^{18}$ eV are of Galactic origin, the particles being the ones which are found in the composition of the stellar winds of stars that explode as supernova into the interstellar medium (ISM) or into their winds. These particles are accelerated in the supernova shock. In order to obtain the observed cosmic ray spectrum it is necessary to take into account the diffusive losses in the Galaxy (which are making the energy spectrum more steeper). Another modification of the source spectrum is due to the fragmentation (spallation) of the cosmic ray particles, after their collision with the ISM atoms. In this paper we are proving that some particles are injected in the supernova shock one or two time ionized, and, also, that the contribution of massive stars ($30 M_{\odot}\leq M\leq 50 M_{\odot}$) accelerated particles to cosmic rays (where the winds are highly enriched in heavy elements) is 1:2 for elements with $Z\geq 6$. Another goal of this paper is to check if the particles are injected with the same velocity, energy or momentum.
\end{abstract}


\section{Introduction}

Cosmic rays are particles, which have been accelerated in the Galaxy or into the extragalactic
space. They come as electrons, protons, heavier nuclei, and their antiparticles. Into
this radiation were detected almost all the elements of the periodic table as well as their isotopes.
Only those particles are missing, that decay with a lifetime generally too short to be
observed on Earth, like neutrons and many radioactive isotopes. The energy spectrum reaches
from some MeV to $10^{20}$ eV/particle, the highest energy of any known radiation. From the
lower end, up to the highest energies, the flux decreases by about 30 orders of magnitude. In
consideration of such a large energy range, different detection techniques are necessary to
explore the cosmic radiation. Methods used at medium and high energy ($<10^{10}$ eV) require
detectors carried on balloons, rockets or satellites and, for studying the charged component
of the cosmic radiation, this technique has been extended up to several $10^{14}$ eV/particle.
At still higher energies the observations are done with detectors located deep underground
and with air shower arrays, which cover the ultra high energy (UHE; $>10^{14}$ eV) and extremely
high energy (EHE) regions ($>10^{18}$ eV). Whereas the lower energy experiments at the top
of the atmosphere measure the primary nuclei directly, the ground-based arrays only detect
the air showers generated by the incoming particles.

Up to now, the origin and acceleration processes of cosmic rays are still subjects of debate. As main sources of cosmic rays are assumed
supernovae, quasars, active galactic nuclei (AGN) and Gamma Ray Bursts (GRB). Because they are deflected in the
interstellar magnetic fields, the charged particles of the cosmic ray radiation appear nearly isotopic on earth, only the neutral constituents such as photons and
neutrinos providing direct information on their acceleration site.

The energy content of the cosmic ray radiation is about 1 eV/cm$^{3}$ and, therefore, comparable with other forms
of energy in our galaxy, like magnetic fields ($\sim$ 1 eV/cm$^{3}$) or starlight 
($\sim$ 0.44 eV/cm$^{3}$).

Recently were proposed three source sites for the observed cosmic rays \cite{wiebel}:

\begin{quote}
\begin{enumerate}
\item Supernova explosions into the interstellar medium, or ISM-SNe. This component produces
mostly hydrogen and the observed energetic electrons up to about 30 GeV, and dominates the
{\it all particle} flux up to $10^{4}$ GeV.
\item Supernova explosion into predecessor stellar wind, or wind-SNe. This component produces
the observed electrons above 30 GeV, helium and most heavier elements already from
GeV particle energies. The component extends ultimately to several EeV. Since the winds of massive stars are enriched late in their life,
this component shows a heavy element abundance which is strongly increased over the ISM.
\item The radio galaxies produce a contribution which dominates beyond about 3 EeV,
and consist mostly of hydrogen and helium, with only little addition of heavy elements below
50 EeV. At this energy the interaction with the microwave background cuts off the contribution
from distant extragalactic sources: the Greisen-Zatsepin-Kuzmin (GZK) cutoff. There are a
small number of events which appear to be beyond this energy, and whether they fit into such
a picture is open at present.
\end{enumerate}
\end{quote}

The theory was originally proposed in Biermann (1993, paper CR I) and in Rachen \& Biermann
(1993, paper UHE CR I). Various tests were performed in Biermann \& Cassinelli 1993, paper
CR II; Biermann \& Strom 1993, paper CR III; Stanev et al. 1993, paper CR IV etc.

\section{Cosmic Ray Acceleration}

Cosmic rays with energies up to 100 TeV are thought to arise predominantly through shock
acceleration by supernova remnants (SNR) in our Galaxy. A fraction of the accelerated cosmic rays
should interact within the supernova remnant and produce gamma-rays. Recent
observations above 100 MeV (the EGRET instrument on the Compton Gamma Ray Observatory)
have found gamma ray signals associated with at least two supernova remnants - IC 443 and
$\gamma$ Cygni. However, it is possible that the gamma ray emission from $\gamma$ Cygni to be
associated with a pulsar within the remnant rather than the remnant itself. Further evidence
for acceleration in SNR comes from the recent ASCA observation of non-thermal X-ray emission
from SN 1006. Reynolds \cite{reynolds} and Mastichiadis \cite{mastichiadis} interpret the latter as synchrotron emission
by electrons accelerated in the remnant up to energies as high as 100 TeV, although Donea and
Biermann \cite{donea} suggest it may be bremsstrahlung from much lower energy electrons.

Acceleration to somewhat higher energies than 100 TeV may be possible, but probably not high
enough to explain the smooth extension of the spectrum to 1 EeV. Several explanations for the
origin of the cosmic rays in this range have been suggested: reacceleration of the supernova
component while still inside the remnant; by several supernovae exploding into a region
evacuated by a pre-supernova star; by shock acceleration inside the strong winds of hot
stars or of groups of hot stars. At 5 EeV the spectral slope changes, and there is evidence
for a lightening in composition. Is likely that this marks a change from galactic cosmic
rays to extragalactic cosmic rays.

For stochastic particle acceleration by electric fields induced by motion of magnetic fields
{\it B}, the rate of energy gain for relativistic particles with charge {\it Ze} can be written
(in SI units):

$$
\left(\frac{dE}{dt}\right)_{acc}=\xi Zec^{2}B\; ,
$$
\noindent
where $\xi<$ 1 depends on the acceleration mechanism.

The second order Fermi acceleration (Fermi's original theory) can be modified in the context
of supernova (SN) shocks, or other strong astrophysical shocks, into the more efficient first order
Fermi mechanism \cite{protheroe}.

\section{Propagation of Galactic Cosmic Rays}

\subsection{The FIP Factor Correction}

The elemental and isotopic mass fractions of cosmic rays, when traced back to their sources,
resemble general mass fractions (or with the general abundances - GA) of elements \cite{silb}, but display
also some significant differences. The general mass fractions are based mainly on measurements
of solar spectra and in carbonaceous chondrite meteorites. For the elements H, He and N the
difference between CR and GA are factors of 20-30 (underabundant in cosmic rays). The
elements O, S, and Ar are underabundant in CR by a factor of about 4 and the nuclide $^{20}$Ne
by a factor of about 6. The elements C and Zn and the nuclide $^{22}$Ne are underabundant by a
factor of about 2.

The cosmic ray source abundances of elements and isotopes are deduced from the abundances of
cosmic rays near Earth, applying propagation corrections for the secondary components produced by 
{\bf nuclear spallation} in the interstellar gas.

The cosmic ray source composition resembles that of normal stars like the Sun,
modified by the photosphere, chromosphere to corona particle escape mechanism that result in
a diminished element abundance for the elements whose first ionization potential (FIP) exceeds about 10 eV \cite{silb}. We note that FIP provides a partial organizing principle of the ratios of cosmic-ray source abundances relative to general abundances. The value of about 10 eV implies a first-stage injection temperature
(at stellar photospheres) of $\sim 10^{4}$ K (about 1 eV), with easier escape for the charged
ones. When these particles become coronal particles, they are boosted up to energies near 1
keV. Some of the coronal particles are boosted up in energy, becoming flare particles, with
energies of the order of 1 MeV. A further stage of injection (stellar flare particles near the
shock waves of supernova remnants) at energies near 1 MeV is plausible, as relative ionization
loss effects on particle range and on the composition cancels due to the effective charges of
atoms near energies of 1 MeV/nucleon.

It is to be noted that the elements (and isotopes of Ne) with FIP $>10$ eV have a very wide
spread, and that other organizing principles in addition to FIP are needed. Since cosmic rays' $^{22}$Ne,
C and O contain contributions from Wolf-Rayet stars, and Ne could be affected
by suppression of light elements, it is useful to adopt the elements S and Ar for the FIP
correction. A value of $I=0.27$ for FIP $\geq 10.4$ eV is adopted, and $I\equiv 1$ for
FIP $\leq 8.4$ eV. Zn, with FIP $=9.4$ eV, is in the middle of the transition
region.

Other forms have been proposed for the FIP correction, a linear fit
being still unsatisfactory (as seen in \cite{silb}) especially for N and $^{22}$Ne.

\subsection{Contribution of Wolf-Rayet Stars to the Composition of Cosmic Rays}

A second difference between the cosmic ray source composition and the general abundance is
the enhancement of $^{22}$Ne, $^{25}$Mg and $^{26}$Mg. Prantzos et al. \cite{prantzos1} showed
that the abundance ratio C/O in cosmic rays (about twice the solar one) can be explained in
terms of Wolf-Rayet star contribution to carbon. The winds of Wolf-Rayet stars are energetic,
close to 0.1 MeV/nucleon, relatively close to the energies of flare particles, and so, the
acceleration of these wind particles is possible. The Wolf-Rayet stars go through two phases:
WN, when the nitrogen produced in CNO hydrogen-burning cycle is abundant at the stellar surface, and WC,
when the carbon, produced in helium burning, is abundant at the surface, and $^{14}$N burns into
$^{22}$Ne \cite{silb}.

\subsection{Spallation Correction}

The spallation is one of the processes having the most important role
in the modification of the abundance of cosmic rays in their transportation
through ISM.

The basic features of target fragmentation, sometimes called ``spallation'', are very
well understood: heavy fragments arise from peripheral collisions of heavy ions or
relativistic protons with the target nucleus. These so called ``spectators'' of the reaction
are excited primary fragments which then decay into the final fragments by a sequence of
evaporation steps.

The spectrum of the residual nuclei seems to be determined to large extent, but not fully, by
the evaporation process.

The difference between the spallation in high-energy physics in accelerators and in
astrophysics is that, in the first case, the targets are the nuclei and the projectiles are the
high-energy protons, while in the second case the targets are the ISM protons (in first
approximation) and the projectiles are the nuclei that form the cosmic radiation. So,
everything depends just on which particle is putted into the reference frame. 
About 10\% by number of the interstellar gas is helium. Hence, about 20\% of cosmic ray generated nuclear
spallation nuclear products are formed in nucleus-helium interactions. In the analysis of
cosmic ray interactions with atmospheric nuclei (mainly nitrogen and oxygen) an accurate
knowledge of nucleus-nucleus interactions becomes essential \cite{tsao}.

High-energy protons cause many different nuclear reactions and, in principle, all these nuclear
processes have to be taken into account for a proper description of the total process. Possible
outgoing particles are, for example, all light particles, gamma rays and (above incident
energies of 150 MeV) pions. Furthermore, high-energy fission may occur and outgoing particles
from the proton bombardment stage further reactions. Clearly, a wide spectrum of reaction
products will be formed. Because of this diversity, the nuclear data needs for
accelerator-based transmutation have been categorized in the following classes:

\begin{enumerate}
\item {\it Proton-induced reaction data}

High-energy proton reactions represent the primary step of the total transmutation or
spallation process, and the importance of nuclear data associated with these reactions is
obvious. Also, secondary protons with lower energy than the bombarding energy are involved.
Needed data include total, elastic, inelastic, spallation cross sections, fission cross
sections and reaction cross sections for outgoing light particles (p, n, d, t, $\alpha$,
$\gamma$, $\pi^{+}$, $\pi^{0}$, $\pi^{-}$) and fission products. In particular, total neutron
and proton production cross sections are relevant. Energy-angle distributions are mainly
required for outgoing high-energy neutrons and protons. Cross-sections for outgoing protons
or neutrons constitute the most important data. For the method based on intense neutron fluxes,
(p, xn) data have highest priority, but also the total proton, neutron and spallation yields
are required.

\item {\it Neutron-induced reaction data}

After the interaction of a high-energy proton with a target nucleus, high-energy neutrons which
have been knocked out by the intranuclear cascade will induce secondary nuclear reactions and/or fission products in
the actinides and in the spallation target. The same
data as needed for proton-induced reactions are required for these high-energy neutrons. The
highly excited residual nucleus that is left after the intranuclear cascade also evaporates a
large number of relatively low-energy neutrons. The low-energy reactions that are induced
by these outgoing neutrons have been subject to extensive evaluations. The additional
neutron-induced data that are required are therefore mainly those for energies between 20 and
1500 MeV.

\item {\it Other nuclear reaction data}
\end{enumerate}

The spallation occurs when the supernova shock travels through the stellar wind and then hits
the surrounding molecular shell of dense gas. Then, there are two ways to consider the spallation: 
first, in the approximation that the spallation is steady, and second, that the
spallation takes place in an expanding medium with a time scale of expansion comparable
to the time scale of the spallation.

\section{The Initial Mass Function (IMF)}

The IMF is assumed to be a time invariant mass spectrum with a power law of the form:
$$
n(m)dm\propto m^{-(1+x)}dm
$$
for which {\it m} is the mass of a main sequence (MS) star in units of $M_{\odot}$, $m_{l}\leq m\leq m_{u}$ 
and {\it n(m)dm} - the number of stars in the mass
interval {\it m} and {\it m+dm}; $m_{l}$ is a lower mass limit of stars, and $m_{u}$ is an
upper limit of stars. Normally, the IMF is derived from the observed present-day mass function
(PDMF) in the solar neighborhood, which is assumed to be independent of time. The derivation
of the IMF from PDMF is difficult, involving assumptions about the star formation rate during
the lifetime of the Galaxy. For stars with lifetimes longer than the age of the Galaxy ($m\leq
1 M_{\odot}$), IMF is derived by assuming an average star formation rate in the past, whereas
for stars with lifetimes negligible relative to the age of the Galaxy ($m\geq 2 M_{\odot}$),
the IMF is derived by assuming a present-time formation rate and taking into account the
stellar lifetimes ($\tau (m)$). Given the uncertainties in both theory and observation, the IMF
variations can be parameterized, and the proposed IMF can be tested by means of a detailed
chemical evolution model. This method has been adopted in many cases. For example, globular clusters 
are expected to have a IMF similar to the Miller-Scalo form, with a typical lower mass limit
of 0.1 $M_{\odot}$. The Miller-Scalo IMF can be well
approximated by a half-Gaussian distribution in $\log m$ :
$$
\phi (\log m)=C_{0}\exp\left[-C_{1}\left(\log m - C_{2}\right)^{2}\right]\; ,
$$
where $C_{0}=66.2$, $C_{1}=1.15$, $C_{2}=-0.88$ \cite{lu}.

\section{Chemical Yields for MS and post-MS Stars}

To have acceleration of wind particles in supernova shocks, first it is necessary to have a
star massive enough to make possible the supernova explosion and, also, in their
pre-supernova stage, strong winds to enrich the circumstellar medium with heavy elements.

In the Hertzsprung-Russell diagram, stars along the main sequence need to be considered in four
separate zero age mass ranges:

\begin{itemize}
\item Stars below about 8 $M_{\odot}$ do not explode as supernovae;

\item Stars from about 8 to about 15 $M_{\odot}$ explode as supernovae, but do not have a
strong stellar wind, and so explode into the interstellar medium. In pre-supernova stage evolve
as Red Supergiants (RSG);

\item Stars from about 15 to 30 $M_{\odot}$ have a substantial wind. This wind is enriched
only in helium. The chemical composition of the wind at the time of explosion is approximately
50\% in He and 50\% in H. The mass in the shell of wind-swept material is moderate. In
pre-supernova stage evolve as RSG;

\item Stars from about 30 $M_{\odot}$ have strong winds. This wind is enriched in heavy
elements and has little hydrogen left. The mass in the shell of wind-swept material is large.
In pre-supernova stage evolve as Wolf-Rayet (WR) stars or as ``luminous blue variables'' (LBV) stars.
\end{itemize}

The evolution of massive stars depends strongly on mass loss by stellar winds. The most
illustrative example is given by the fact that massive stars with initial masses above 25
$M_{\odot}$ are likely to be left with only about 5 $M_{\odot}$ at the end of their Wolf-Rayet
stage. Wolf-Rayet stars represent a late evolutionary stage of stars with masses of 25-50 
$M_{\odot}$ which occurs as a result of the loss of all or at least the major part
of their hydrogen-rich envelope. Although WR stars are among the most luminous stellar objects
in the sky and contribute substantially to the chemical and dynamical evolution of galaxies,
even their basic properties, e.g. their radii or their luminosities, are still not well known
today. This arises mainly from the fact that WR stars are obscured by their own dense stellar
wind, the origin of which is not yet understood and which renders the modeling of WR stars and
their atmospheres difficult \cite{heger}.

Through frequently modeled, the hydrogen and helium burning stages of massive stellar evolution
are greatly affected by uncertain physics: convection, mass loss and the cross section for the
$^{12}$C$(\alpha, \gamma)^{16}$O reaction. The extent of convective penetration into regions
which are stable according to local criteria (``convective overshooting'') is especially
uncertain. Observational arguments in favor of a moderate amount of overshooting is, for
example, the main-sequence widening, but those arguments are indirect, and other processes
such as rotationally induced mixing or improvements in radiative opacities might satisfy the
same observational constraints.

A second uncertainty concerns the efficiency of molecular weight gradients in preventing
convective mixing (``semiconvection''). Two extreme assumptions are frequently seen in the
literature:

\begin{enumerate}
\item that molecular weight barriers may be neglected altogether (``Schwarzschild criterion'' for convection);
\item that molecular weight barriers prevent almost any mixing (``Ledoux criterion'' for
convection).
\end{enumerate}

In between these extremes, semiconvection describes the rate of mixing in regions having
superadiabatic temperature gradients, but stabilized by finite molecular weight gradients.

The physics of mass loss is not sufficiently well understood for almost any evolutionary phase
of massive stars to make a reliable quantitative prediction. Progress has been achieved
recently in the theory of radiatively driven winds, which is applicable to the hot stages
($T_{eff}\geq 15,000$ K) of massive hydrogen-rich stars, but this still involves considerable
uncertainty and does not agree with observed mass-loss rates to better than a factor of 2. Much
more uncertain are the mass-loss rates for red supergiants. Only approximate fits to the
observations exist and a factor of 10 uncertainty may not be an overestimate. Finally,
mass-loss rates for LBVs and for the various types of WR stars are also uncertain.

\section{Mass Loss and its Effects on the Chemical Yields}

Mass loss by stellar winds is a dominant effect in the evolution of stars with an initial mass
$M\geq 20 M_{\odot}$. The standard picture for the pre-supernova structure of a massive star
(at least near solar metallicity) is that of a red supergiant which contains a hydrogen-rich
envelope of many solar masses at the time of explosion. However, the inclusion of mass loss
allows several alternative structures, all of which presumably have observational counterparts
as supernovae. There may be red supergiants with a very small hydrogen-rich envelope,
hydrogen-rich WR stars, helium stars (WR stars with a surface helium mass fraction $Y_{S}=1-Z$, with {\it Z} as initial metallicity), or WR stars of WC or WO type, which contain large amounts of He, C and O (and no hydrogen) at their surface.

The star may become a bare core, then enter the WR stage and follow all or part of the sequence
WNL (L for late), WNE (E for early), WCL, WCE and WO. This sequence corresponds to a
progression in the exposition of the nuclear products: CNO equilibrium with H present (WNL),
CNO equilibrium without H (WNE), early visibility of the products of the $3\alpha$ reaction
(WCL) and the growing of the (C+O)/He ratio in the WCE and WO stars. The comparison of the
observed and predicted abundance ratios confirms the validity of the views on the CNO cycle
and He-burning.

The evolution of massive stars is highly dependent on the initial stellar metallicity {\it Z}
through
the effects of mass loss. The mass loss depends on metallicity because the radiation pressure
on the envelope depends on the abundance of metallic ions. The mass loss $\dot{M}$ in O- and
B-type stars varies like $\dot{M}\sim Z^{\alpha}$, with a value of $\alpha$ between 0.5 and 1.
When the star enters the WNE stage and after the disappearance of its hydrogen, a very
important effect is that its mass loss rates essentially behave like $\dot{M}\sim M^{2.5}$,
probably with no dependence on the initial {\it Z}. Consequently, the $\dot{M}$-rates are very
large
at the beginning of the WNE stage (up to $10^{-2}M_{\odot}$/year for $M\simeq 50 M_{\odot}$)
and they then decline very rapidly as the stellar mass is decreasing. This produces a
remarkable mass convergence (for example, all stars at solar metallicity with initial $M\geq
25 M_{\odot}$ reach the pre-SN stage with a final mass of about $5 M_{\odot}$). The metallicity
dependence of the mass loss rates has been included in various grids and models. As a result
the relation between the final and initial masses is very different according to the initial
{\it Z}. At solar {\it Z} one has the above-mentioned mass convergence, while at very low {\it
Z} the final masses are very close to the initial ones.

The initial metallicity of the star affects not only mass loss, but also the outcome of
nucleosynthesis. For metallicities $Z<Z_{\odot}/20$ mass loss has presumably a negligible
effect on the yields of stars of all masses. During H-burning, the initial CNO transforms to
$^{14}$N, and part of the latter nucleus turns into $^{22}$Ne during He-burning (through
$\alpha$ captures and one $\beta$ decay). $^{12}$C, $^{14}$N and $^{16}$O all have equal
numbers of neutrons and protons but not $^{22}$Ne (10 protons and 12 neutrons). This surplus of
neutrons (increasing with initial metallicity) affects the products of subsequent burning
stages and, in particular, of explosive burning, favoring the production of odd nuclei
(``odd-even'' effect) \cite{prantzos}. \\

{\bf Mass Loss Rates}

\begin{itemize}
\item The mass loss rate employed during helium burning (up to carbon ignition) is $\dot{M} =
-kM^{2.5}$ (with {\it M} in $M_{\odot}$ and $\dot{M}$ in $M_{\odot}$/year) and $k=6\times
10^{-8} M_{\odot}$/year, so long as the carbon surface mass fraction does not exceed 0.02, and 
$k=10^{-7} M_{\odot}$/year afterward \cite{langer}. This is also the mass loss rate taken for hydrogen-free 
WR phase, but with $k=6\times 10^{-8} M_{\odot}$/year \cite{segura}.
\item For WR stars with a nonvanishing surface hydrogen abundance (WNL stars) the mass loss
might be assumed constant: $3\times 10^{-5} M_{\odot}$/year for $M<25 M_{\odot}$, $5\times
10^{-5} M_{\odot}$/year for $25 M_{\odot}<M<35 M_{\odot}$ and $8\times 10^{-5} M_{\odot}$/year
for $M>35 M_{\odot}$ \cite{schmutz}.
\item For hydrogenless WR stars a mass loss rate as follows can be adopted:

$$
\dot{M}=-k\left(\frac{M}{M_{\odot}}\right)^{\alpha}\; ,
$$

with $\alpha=2.6$, and $k=5\times 10^{-8} M_{\odot}$/year for hydrogenless WN stars (WNE stars)
and $k=10^{-7} M_{\odot}$/year for WC/WO stars \cite{langer1}.
\end{itemize}

\section{Effects of Rotation on the Surface Abundances and Chemical Yields}

For non-rotating stars, the surface enrichment in H and N occurs when the star reaches the RSG
phase. There, CNO elements are dredged-up by deep convection. For rotating stars, N-excesses
occur already during the MS phase, for solar metallicity $Z=0.02$, the predicted excesses
amount to factors 3 and 4 for initial $v_{rot}=$ 200 and 300 km/s respectively. At lower
metallicity, the N-enrichment during the MS phase is smaller, due to the lower mass loss.
However, there is a very large increase (up to a factor of $\sim 10$) for late B-type
supergiants, because at low {\it Z} the star spends a lot of time in the blue phase and mixing
processes have time to work. These predictions are in agreement with the observed excesses for
galactic B- and A-type supergiants. Also, the very large excesses observed for A-type
supergiants in the Small Magellanic Cloud are remarkably well accounted for.

The chemical yields are strongly modified by rotation. The larger He-cores obtained in rotating
models at core collapse imply larger productions of He and other $\alpha$-nuclei elements. This
is the most important effect of rotation on the chemical yields. In addition, by enhancing the
mass loss rates and making the formation of WR stars easier, rotation favors the enrichment of
the ISM by stellar winds.

The rotational diffusion during the H-burning phase enriches the outer layers in CNO processed
elements. Some $^{14}$N is extracted from the core and saved from further destruction. The same
can be said for $^{17}$O and $^{26}$Al, a radioisotope with a half-life of 0.72 Myr. The mixing
in the envelope of rotating stars also leads to a faster depletion of the temperature sensitive
light isotopes, for instance lithium and boron \cite{fliegner}. The supergiants at low {\it Z} in the
range of 9 to 20 $M_{\odot}$ seem to be good candidates for the production of primary nitrogen
\cite{maeder}.

\section{Our Model}

After a discussion with P.L. Biermann we decided that, for computing the mass fractions of different elements in the winds of massive stars, it is useful to write:

\begin{equation}
\label{eq_a}
\displaystyle \bar{X}_{i}=\frac{\int_{M_{i}}^{M_{f}}{X_{i}(M)\;\Phi{(M)}\;\dot{M}(X_{i})\;dM}}{\int_{M_{i}}^{M_{f}}{\Phi{(M)}\;\dot{M}(X_{i})\;dM}}\; ,
\end{equation}
\noindent
where $X_{i}(M)$ is the mass fraction of element {\it i} (see Table \ref{abund}) in the stellar wind ejecta, for one
computed stellar model (with given semiconvection and metallicity) and for one
initial stellar mass.

\begin{table*}
\begin{center}
\begin{footnotesize}
\caption{Surface mass fractions of various isotopes in stellar evolution models in the initial
mass range $15 M_{\odot}\leq M_{init}\leq 50 M_{\odot}$ at the pre-SN stage. The pre-SN
configuration is also indicated, where RSG means red supergiant and WC stands for Wolf-Rayet
star of the carbon sequence. The last column gives the initial mass fractions used in the stellar
evolution calculations \cite{langer2}.}
\begin{tabular}{c c c c c c c c} \hline \hline
 & $15M_{\odot}$ & $20M_{\odot}$ & $25M_{\odot}$ & $30M_{\odot}$ & $40M_{\odot}$ &
$50M_{\odot}$ & initial \\
isotop & RSG & RSG & RSG & RSG & WC & WC & \\
\hline
$^{1}$H & $6.75\cdot 10^{-1}$ & $6.48\cdot 10^{-1}$ & $6.31\cdot 10^{-1}$ & $6.16\cdot 10^{-1}$
& 0.0 & 0.0 & $7.00\cdot 10^{-1}$ \\
$^{4}$He & $3.06\cdot 10^{-1}$ & $3.32\cdot 10^{-1}$ & $3.49\cdot 10^{-1}$ & $3.65\cdot 10^{-1}
$ & $7.22\cdot 10^{-1}$ & $1.49\cdot 10^{-1}$ & $2.80\cdot 10^{-1}$ \\
$^{12}$C & $2.40\cdot 10^{-3}$ & $2.00\cdot 10^{-3}$ & $2.05\cdot 10^{-3}$ & $2.01\cdot 10^{-3}
$ & $2.07\cdot 10^{-1}$ & $4.94\cdot 10^{-1}$ & $3.48\cdot 10^{-3}$ \\
$^{13}$C & $1.04\cdot 10^{-4}$ & $2.08\cdot 10^{-4}$ & $1.35\cdot 10^{-4}$ & $1.39\cdot 10^{-4}
$ & $4.70\cdot 10^{-7}$ & 0.0 & $3.87\cdot 10^{-5}$ \\
$^{14}$N & $3.04\cdot 10^{-3}$ & $4.17\cdot 10^{-3}$ & $4.42\cdot 10^{-3}$ & $4.83\cdot 10^{-3}
$ & $4.51\cdot 10^{-3}$ & 0.0 & $1.03\cdot 10^{-3}$ \\
$^{15}$N & $2.27\cdot 10^{-6}$ & $2.05\cdot 10^{-6}$ & $1.69\cdot 10^{-6}$ & $1.62\cdot 10^{-6}
$ & 0.0 & 0.0 & $3.77\cdot 10^{-6}$ \\
$^{16}$O & $9.09\cdot 10^{-3}$ & $8.21\cdot 10^{-3}$ & $7.85\cdot 10^{-3}$ & $7.42\cdot 10^{-3}
$ & $4.81\cdot 10^{-2}$ & $3.32\cdot 10^{-1}$ & $9.98\cdot 10^{-3}$ \\
$^{17}$O & $5.69\cdot 10^{-5}$ & $6.04\cdot 10^{-5}$ & $6.46\cdot 10^{-5}$ & $6.95\cdot 10^{-5}
$ & $8.33\cdot 10^{-6}$ & 0.0 & $3.80\cdot 10^{-6}$ \\
$^{18}$O & $1.18\cdot 10^{-5}$ & $3.33\cdot 10^{-6}$ & $8.67\cdot 10^{-6}$ & $8.22\cdot 10^{-6}
$ & $8.92\cdot 10^{-4}$ & 0.0 & $2.00\cdot 10^{-5}$ \\
$^{19}$F & $3.26\cdot 10^{-7}$ & $3.44\cdot 10^{-7}$ & $2.73\cdot 10^{-7}$ & $2.53\cdot 10^{-7}
$ & $0.03\cdot 10^{-9}$ & $0.05\cdot 10^{-9}$ & $3.74\cdot 10^{-7}$ \\
$^{20}$Ne & $1.70\cdot 10^{-3}$ & $1.70\cdot 10^{-3}$ & $1.70\cdot 10^{-3}$ & $1.70\cdot 
10^{-3}$ & $1.68\cdot 10^{-3}$ & $1.66\cdot 10^{-3}$ & $1.70\cdot 10^{-3}$ \\
$^{21}$Ne & $3.93\cdot 10^{-6}$ & $3.90\cdot 10^{-6}$ & $3.28\cdot 10^{-6}$ & $3.05\cdot 
10^{-6}$ & $1.38\cdot 10^{-5}$ & $3.61\cdot 10^{-5}$ & $4.14\cdot 10^{-6}$ \\
$^{22}$Ne & $1.10\cdot 10^{-4}$ & $1.00\cdot 10^{-4}$ & $9.77\cdot 10^{-5}$ & $9.43\cdot 
10^{-5}$ & $1.21\cdot 10^{-2}$ & $1.84\cdot 10^{-2}$ & $1.24\cdot 10^{-4}$ \\
$^{23}$Na & $5.10\cdot 10^{-5}$ & $5.56\cdot 10^{-5}$ & $6.53\cdot 10^{-5}$ & $6.98\cdot 
10^{-5}$ & $1.78\cdot 10^{-4}$ & $1.76\cdot 10^{-4}$ & $3.46\cdot 10^{-5}$ \\
$^{24}$Mg & $5.38\cdot 10^{-4}$ & $5.38\cdot 10^{-4}$ & $5.38\cdot 10^{-4}$ & $5.38\cdot 
10^{-4}$ & $4.11\cdot 10^{-4}$ & $2.26\cdot 10^{-4}$ & $5.38\cdot 10^{-4}$ \\
$^{25}$Mg & $6.52\cdot 10^{-5}$ & $6.22\cdot 10^{-5}$ & $5.92\cdot 10^{-5}$ & $5.70\cdot 
10^{-5}$ & $2.12\cdot 10^{-4}$ & $9.34\cdot 10^{-4}$ & $6.81\cdot 10^{-5}$ \\
$^{26}$Mg & $7.86\cdot 10^{-5}$ & $8.08\cdot 10^{-5}$ & $8.34\cdot 10^{-5}$ & $8.52\cdot 
10^{-5}$ & $3.90\cdot 10^{-4}$ & $2.03\cdot 10^{-3}$ & $7.50\cdot 10^{-5}$ \\
$^{26}$Al & $0.25\cdot 10^{-9}$ & $5.22\cdot 10^{-9}$ & $9.07\cdot 10^{-8}$ & $2.99\cdot 
10^{-7}$ & $4.09\cdot 10^{-7}$ & $3.60\cdot 10^{-7}$ & 0.0 \\
$^{27}$Al & $6.05\cdot 10^{-5}$ & $6.06\cdot 10^{-5}$ & $6.07\cdot 10^{-5}$ & $6.10\cdot 
10^{-5}$ & $7.75\cdot 10^{-5}$ & $8.02\cdot 10^{-5}$ & $6.00\cdot 10^{-5}$ \\
$^{28}$Si & $6.80\cdot 10^{-4}$ & $6.80\cdot 10^{-4}$ & $6.80\cdot 10^{-4}$ & $6.80\cdot 
10^{-4}$ & $5.59\cdot 10^{-4}$ & $3.63\cdot 10^{-4}$ & $6.80\cdot 10^{-4}$ \\
$^{29}$Si & $3.44\cdot 10^{-5}$ & $3.44\cdot 10^{-5}$ & $3.44\cdot 10^{-5}$ & $3.44\cdot 
10^{-5}$ & $1.06\cdot 10^{-4}$ & $1.49\cdot 10^{-4}$ & $3.44\cdot 10^{-5}$ \\
$^{30}$Si & $2.28\cdot 10^{-5}$ & $2.28\cdot 10^{-5}$ & $2.28\cdot 10^{-5}$ & $2.28\cdot 
10^{-5}$ & $7.86\cdot 10^{-5}$ & $2.43\cdot 10^{-4}$ & $2.28\cdot 10^{-5}$ \\
$^{56}$Fe & $1.32\cdot 10^{-3}$ & $1.32\cdot 10^{-3}$ & $1.32\cdot 10^{-3}$ & $1.32\cdot 
10^{-3}$ & $1.32\cdot 10^{-3}$ & $1.32\cdot 10^{-3}$ & $1.32\cdot 10^{-3}$ \\

\hline
\end{tabular}
\label{abund}
\end{footnotesize}
\end{center}
\end{table*}

Here $\Phi(M)$ is the Miller-Scalo IMF:

$$
\Phi(\log{M})=C_{0}\exp{\left[-C_{1}\left(\log{M}-C_{2}\right)^{2}\right]} \; ,
$$
\noindent
where $C_{0}=66.2$, $C_{1}=1.15$, $C_{2}=-0.88$. $M$ is the mass of MS star
in units of $M_{\odot}$.

In equation (\ref{eq_a}), $M_{i}$ and $M_{f}$ are the initial and final masses
between which the stars exploding as supernova enrich the ISM.
For obtaining the contribution of all the masses from H-R diagram one must make a
summation after all those intervals between 15 and 50 solar masses (so, for the
stars that are having strong winds).

The mass range is split in intervals because in a representation
$\log{X_{i}}=f(\log{M})$:

1) There is no unique function to describe the evolution (in logarithmic
scale) of the mass fractions with the mass. In the first approximation, 
this evolution can be described by a step function.
The first reason for this behavior is that stars from different mass ranges have different wind mass loss 
and so, different contributions to the enrichment of ISM. The second reason is that the mechanism through which the 
wind is driven changes at 25-30 $M_{\odot}$;

2) The composition in the wind between 15 and 25 $M_{\odot}$ is dominated by He, and
between 25 and 50 $M_{\odot}$ by C and O.

The mass loss $\dot{M}_{i}$ for the element $i$ (in $M_{\odot}$) is:

$$
\dot{M}(X_{i})=\int X_{i,\;surf}(t,M)\;\dot{M}(t,M)\;dt \; .
$$

Assuming that the mass loss is constant in time, the above equation will become:

$$
\dot{M}(X_{i})=X_{i,\;surf}(M)\; \dot{M}(M)=X_{i}(M)\; \dot{M}(M)\; .
$$

For practical reasons it is useful to put all these terms as function of $\ln{M}$ instead of $M$. Then, the equation (\ref{eq_a}) can be written:

\begin{equation}
\bar{X}_{i}=\frac{\int_{\ln{M_{i}}}^{\ln{M_{f}}}{X_{i}^{2}(M)\;\Phi(M)\;M\;\dot{M}(M)\;dM}}{\int_{\ln{M_{i}}}^{\ln{M_{f}}}{X_{i}(M)\;\Phi(M)\;M\;\dot{M}(M)\;dM}} \; ,
\label{model}
\end{equation}
\noindent
where we introduced the expression for $\dot{M}(X_{i})$ and we assumed that $X_{i}\approx X_{i,\;surf}$.

We will have $X_{i}(\ln{M})=\exp({\cal F})$, where ${\cal F}=a\ln{M}+b$ is the
fitting function for $\ln{X_{i}}=f(\ln{M})$ for the domain in which the integral is defined.

\begin{eqnarray}
\Phi(\ln{M})&=&\exp{\left[\ln C_{0}-C_{1}(0.434\times \ln{M}-C_{2})^{2}\right]}\; .
\end{eqnarray}

$\dot{M}=\exp({{\cal G}})$, where ${\cal G}=\exp({\ln{k}+\alpha\ln{M}})$,
with $k$ and $\alpha$ given into the section 6. ${\cal G}$ can be taken as the logarithm of some numerical value of the assumed stellar evolution model, for the stellar
mass range in which is done the integration and, most important, for the considered element.

Introducing all this functions under the integral we can compute the mass fractions in the wind.

In order to obtain the mass fractions in CR it is necessary to compute the ratios between the
observed cosmic rays and the mass fractions from our proposed model for one example.
Through this we will be able to evidentiate not only the general form, but also and some particular features of our formulas.

Plotting the data of individual nuclei as provided by the direct experiments shows, in
general, a fair correspondence in overlapping energy ranges. The spectra can be described by
simple power laws in energy:

$$
\phi=\phi_{0}E^{-\gamma} \; ,
$$
\noindent
where $\phi$ is the integral differential flux in particles/(m$^{2}$ s sr TeV/nucleus),
$\phi_{0}$ are the absolute flux normalizations and $\gamma$ the spectral index
for various elements. The fitting was done by Wiebel-Sooth \& Biermann. Their results for
elements between H and Ni can be seen in Table \ref{fit}.

\begin{table} [htp]
\centering
\caption{The spectral indices $\gamma$ and absolute flux normalization $\phi_{0}$ for the
various elements with nuclear charge number $Z$. We can see that the spectral index for Hydrogen 
is close to 2.74, and for all the other elements (except Li, Be and B which are spallation products) it is 
close to 2.67 \cite{wiebel}.}
\begin{tabular}{c c c c c} \hline\hline
Element & Z & $\phi_{0}$ [m$^{2}$ s sr TeV/nucleus] $^{-1}$ & $\gamma$ & $\frac{\chi^{2}}{df}$ \\ \hline
H & 1 & $(10.57\pm 0.30)\cdot 10^{-2}$ & $2.76\pm 0.02$ & 0.70 \\
He & 2 & $(6.73\pm 0.20)\cdot 10^{-2}$ & $2.63\pm 0.02$ & 2.10 \\
Li & 3 & $(2.08\pm 0.51)\cdot 10^{-3}$ & $2.54\pm 0.09$ & 0.90 \\
Be & 4 & $(4.74\pm 0.48)\cdot 10^{-4}$ & $2.75\pm 0.04$ & 0.37 \\
B & 5 & $(8.95\pm 0.79)\cdot 10^{-4}$ & $2.95\pm 0.05$ & 0.45 \\
C & 6 & $(1.06\pm 0.01)\cdot 10^{-2}$ & $2.66\pm 0.02$ & 1.42 \\
N & 7 & $(2.35\pm 0.08)\cdot 10^{-3}$ & $2.72\pm 0.05$ & 1.91 \\
O & 8 & $(1.57\pm 0.04)\cdot 10^{-2}$ & $2.68\pm 0.03$ & 1.70 \\
F & 9 & $(3.28\pm 0.48)\cdot 10^{-4}$ & $2.69\pm 0.08$ & 0.47 \\
Ne & 10 & $(4.60\pm 0.10)\cdot 10^{-3}$ & $2.64\pm 0.03$ & 3.14 \\
Na & 11 & $(7.54\pm 0.33)\cdot 10^{-4}$ & $2.66\pm 0.04$ & 0.36 \\
Mg & 12 & $(8.01\pm 0.26)\cdot 10^{-3}$ & $2.64\pm 0.04$ & 0.10 \\
Al & 13 & $(1.15\pm 0.15)\cdot 10^{-3}$ & $2.66\pm 0.04$ & 1.24 \\
Si & 14 & $(7.96\pm 0.15)\cdot 10^{-3}$ & $2.75\pm 0.04$ & 0.10 \\
Mn & 25 & $(1.35\pm 0.14)\cdot 10^{-3}$ & $2.46\pm 0.22$ & 5.38 \\
Fe & 26 & $(1.78\pm 0.18)\cdot 10^{-2}$ & $2.60\pm 0.09$ & 1.81 \\ \hline
\end{tabular}
\label{fit}
\end{table}

In the particular case of, let say, carbon:

\begin{eqnarray}
\label{t}
\phi_{0,C}=\phi_{0,^{12}C}+\phi_{0,^{13}C} \; ,
\end{eqnarray}
\noindent
where:
$$
\gamma_{^{12}C}=\gamma_{^{13}C}=\gamma_{C} \; .
$$

From (\ref{t}):

$$
N_{C}(E)=N_{^{12}C}(E)+N_{^{13}C}(E) \; ,
$$
\noindent
which, in the general case, will behave like ({\it j} is for the isotopes):

$$
N_{sum,\; i}(E)=\sum_{j}{N_{ij}}(E)\; ,
$$
\noindent
with:

$$
N_{sum,\;i}^{observed}(E)\simeq N_{sum,\;i}^{CR}(E)\; .
$$

Then, taking the minimum energy at a given rigidity $p_{i}^{\ast}/Z$ and considering
that is not existing one superior limit for energy at which the particle can be accelerated, the number of particles of one type will be:

$$
N_{i}^{observed}=\int_{E^{\ast}} N_{sum,\;i}^{theor}(E)\; dE=\int_{p_{i}^{\ast}} \phi_{0,i}\left(\frac{pc}{E_{0}}\right)^{-\gamma}d\left(\frac{pc}{E_{0}}\right)=\phi_{0,i}\;\frac{1}{\gamma_{i}-1}\left(\frac{p_{i}^{\ast}c}{E_{0}}\right)^{1-\gamma_{i}}\; .
$$

Now, because the mass fraction:

$$
X_{i}=\frac{N_{i}\;m_{i}}{\sum_{k}{N_{k}\;m_{k}}}\; ,
$$
\noindent
with $k$ = H,..,$i$,...,Fe , the observed cosmic ray mass fractions ratio will be (for carbon):

$$
\frac{X_{C}}{X_{He}}=\frac{N_{C}^{CR}\;m_{C}}{N_{He}^{CR}\;m_{He}}\; ,
$$
\noindent
or, introducing $N_{C}^{CR}$ and $N_{He}^{CR}$ in the above formula:

\begin{eqnarray}
\label{mos}
\displaystyle \frac{X_{C}}{X_{He}}=\frac{\displaystyle \phi_{0,C}\;\frac{1}{\gamma_{C}-1}\left(\frac{p_{C}^{\ast}\;c}{E_{0}}\right)^{1-\gamma_{C}}}{\displaystyle \phi_{0,He}\;\frac{1}{\gamma_{He}-1}\left(\frac{p_{He}^{\ast}\;c}{E_{0}}\right)^{1-\gamma_{He}}}\;\frac{m_{C}}{m_{He}}\; ,
\end{eqnarray}
\noindent
which can be simplified using the rigidity $\displaystyle \frac{p_{C}^{\ast}}{Z_{C}}=\frac{p_{He}^{\ast}}{Z_{He}}$:

\begin{eqnarray}
\label{teoretic}
\frac{X_{C}}{X_{He}}=\frac{\phi_{0,C}}{\phi_{0,He}}\frac{\gamma_{He}-1}{\gamma_{C}-1}\left(\frac{Z_{C}}{Z_{He}}\right)^{1-\gamma_{C}}\left(\frac{p^{\ast}_{He}c}{E_{0}}\right)^{\gamma_{He}-\gamma_{C}}\frac{m_{C}}{m_{He}}\; .
\end{eqnarray}

In equation (\ref{teoretic}), $p_{He}^{\ast 2}=\sqrt{2m_{He}E_{0}}$,
with $E_{0}=100$ MeV the lower energy cutoff.

The cosmic ray ratio between mass fractions will be:

$$
\frac{X_{C}}{X_{He}}=\frac{X_{^{12}C}+X_{^{13}C}}{X_{He}} \; ,
$$
\noindent
with the general form:

$$
\frac{X_{i}}{X_{He}}=\frac{\sum_{j}{X_{ij}}}{X_{He}}\; .
$$

In our model, for obtaining the real CR mass fractions, is necessary to take into account also the FIP correction factors (for ionization losses):

$$
\frac{X_{i}}{X_{He}}=I_{element}\; \frac{\sum_{i}{X_{i}}}{X_{He}} \; .
$$

Now, the mass fraction ratio of any element present in cosmic rays
and He can be computed and compared with the model results for the
same ratios.

\section{Results}

After several tests we saw that the stellar mass loss was just a second order
correction, being possible to neglect it.
Without this, the model mass fraction formula (\ref{model}) becomes:

\begin{equation}
\bar{X}_{i}=\frac{\int_{\ln{M_{i}}}^{\ln{M_{f}}}{X_{i}^{2}(M)\;\Phi(M)\;M\;dM}}{\int_{\ln{M_{i}}}^{\ln{M_{f}}}{X_{i}(M)\;\Phi(M)\;M\;dM}}\; .
\label{masic}
\end{equation}

In our computations we had to consider also that, between two Table \ref{abund} mass values, the mass fractions in the
wind are constant.

First, we took an equal mass fraction contribution of RSG and WR stars to the
production of the observed cosmic radiation. In this case, the obtained values are summarized in Table \ref{mass} 
where, in the second column is the nuclear charge number and in the third the mass number. The
column noted with FIP contains the first ionization potential for the considered element.
In the ninth column we have the needed factor for equality between the model and
cosmic ray abundances, each normalized to the corresponding mass fraction for helium.
$min$ and $max$ are the minimal and, respectively, maximal factors with which we have to multiply the model mass fraction ratios in order 
to take account also of the spallation correction ratios and obtain the experimental value.

\begin{table*}
\begin{center}
\begin{footnotesize}
\caption{Our model data considering that RSG and WR stars are having an equal contribution to the
production of the observed cosmic radiation.}
\begin{tabular}{c c c c c c c c c c c} \hline\hline
elem. & Z & A & $\bar{X}_{i,RSG}^{model}$ & $\bar{X}_{i,WR}^{model}$ & $\bar{X}_{i}^{model}$ &
$\left(\frac{\bar{X}_{i}}{\bar{X}_{He}}\right)^{model}$ & FIP & factor & min & max  \\
& & & & & & & [eV] & model/observed & (for & (for \\
& & & & & & & & & $\gamma_{observed}$) & $\gamma_{observed}$) \\ \hline
H & 1 & 2 & 1.051034 & 0.0 & 1.051034 & 1.364058 & 13.6 & 1/1.42 & 5.33 & - \\
He & 2 & 4 & $6.45098\cdot 10^{-1}$ & $1.31679\cdot 10^{-1}$ & $7.7052\cdot 10^{-1}$ & 1 & 24.4 & - & - & - \\
C & 6 & 12 & $4.22546\cdot 10^{-3}$ & $3.77525\cdot 10^{-2}$ & $4.19782\cdot 10^{-2}$ & 0.05448 & 11.27 & $1/(1.42\cdot 43)$ & 1 &
2 \\
N & 7 & 14 & $7.45885\cdot 10^{-3}$ & $8.22535\cdot 10^{-2}$ & $8.97124\cdot 10^{-2}$ & 0.11643 & 14.52 & 1/43 & 1 & 2 \\
O & 8 & 16 & $1.6265\cdot 10^{-2}$ & $8.7725\cdot 10^{-3}$ & $2.50375\cdot 10^{-2}$ & 0.03249 & 13.56 & 1 & 1 & $\sim 1$ \\
F & 9 & 19 & $5.99832\cdot 10^{-7}$ & $5.47141\cdot 10^{-12}$ & $5.99837\cdot 10^{-7}$ & $7.7848\cdot 10^{-7}$ & 17.4 & $15\div 20$ & 1 & 2 \\
Ne & 10 & 20 & $3.29658\cdot 10^{-3}$ & $3.06398\cdot 10^{-4}$ & $3.60298\cdot 10^{-3}$ & $4.676\cdot 10^{-3}$ & 21.48 & 1 & 1 & $\sim 1$ \\
Na & 11 & 23 & $1.11809\cdot 10^{-4}$ & $3.24637\cdot 10^{-5}$ & $1.44272\cdot 10^{-4}$ & $1.8724\cdot 10^{-4}$ & 5.14 & 1 & 0.97 & 1 \\
Mg & 12 & 24 & $1.04327\cdot 10^{-3}$ & $7.49583\cdot 10^{-5}$ & $1.11823\cdot 10^{-3}$ & $1.4513\cdot 10^{-3}$ & 7.64 & 1 & 0.62 & 1 \\
Al & 13 & 27 & $1.17598\cdot 10^{-4}$ & $1.41344\cdot 10^{-5}$ & $1.31733\cdot 10^{-4}$ & $1.7096\cdot 10^{-4}$ & 5.98 & 1 & 0.55 & 1 \\
Si & 14 & 28 & $1.31863\cdot 10^{-3}$ & $1.01951\cdot 10^{-4}$ & $1.42058\cdot 10^{-3}$ & $1.8436\cdot 10^{-3}$ & 8.15 & 1 & 1 & $\sim 1$ \\
Fe & 26 & 56 & $2.5597\cdot 10^{-3}$ & $2.40742\cdot 10^{-4}$ & $2.80044\cdot 10^{-3}$ & $3.6345\cdot 10^{-3}$ & 7.90 & 20 & 1 & 2 \\ \hline
\end{tabular}
\label{mass}
\end{footnotesize}
\end{center}
\end{table*}

$$
min\times \left(\frac{X_{i}}{X_{He}}\right)^{model}\leq \left(\frac{X_{i}}{X_{He}}\right)^{observed}\leq max\times \left(\frac{X_{i}}{X_{He}}\right)^{model}.
$$

In what follows we tried to prove that, after considering the spallation correction, the factor that differentiate the model
mass fraction ratios from the observed ones (for even-{\it Z} elements) is related to the 
ionization losses (abundance decrease) for elements whose FIP exceeds about 10 eV,
due to the particle escape mechanism from the photosphere and the chromosphere to corona, and, also, to the 
different radiative preacceleration (before the shock injection) of WR star 
wind particle species. Proving that the radiative acceleration plays an important role in the abundance of 
cosmic rays will be an independent study and, because of its complexity, will make the subject of another paper.

Analyzing the data in the Table \ref{mass} we see that the column 4 mass fractions are too high. Second, the columns 10 and 11 factors are not even allowing 
the spallation correction for even-Z elements (see O, Ne,Si), and no FIP correction factor (see column 9) 
is evident for the elements with FIP$\geq 10$ eV.

In this theoretical interpretation we can not count on a correct value for the observed cosmic ray hydrogen because its principal contribution to the interstellar medium enrichment and cosmic ray spectrum doesn't come from stars with $15 M_{\odot}\leq M\leq 50 M_{\odot}$, but from other sources: 
low mass stars (ISM SNe and RSG SNe) and Active Galactic Nuclei (AGN) jets.

So, it is necessary to consider that the WR and RSG stars are having a different contribution to an element abundance in CRs. The formula which describes this can be written:

\begin{eqnarray}
\label{alfa}
\centering
\bar{X}_{i}^{model}=\alpha\; \bar{X}_{i,RSG}^{model}+\left(1-\alpha\right)\bar{X}_{i,WR}^{model}\; ,
\end{eqnarray}
\noindent
where for the previous studied case we had $\alpha=1/2$.

In the same time we need a formula for the radiative preacceleration correction in the
 stellar wind and which, after introduction of the spallative correction, to give an
overabundance in the wind (for Mg, Si and Fe), relative to the observed CRs, situated between 4
and 6 (if we are taking 1 as reference level for {\it CR/surface} abundances for C, O and Ne).

For clarity, for a FIP correction ``hidden'' in the
mass fraction ratio, {\it observed/model} it is the same as saying {\it CR/surface}.

We took first the preacceleration correction as being:

\begin{eqnarray}
\label{prop}
\left(\frac{\bar{X}_{i}}{\bar{X}_{He}}\right)^{observed}\simeq \left(\frac{Z_{i}}{Z_{He}}\right)^{\gamma_{inj}-1}\frac{A_{He}}{A_{i}}\left(\frac{\bar{X}_{i}}{\bar{X}_{He}}\right)^{model}\; .
\end{eqnarray}

In this way:

$$
\left(\frac{\bar{X}_{i}}{\bar{X}_{He}}\right)^{observed}=\left(\frac{Z_{i}}{Z_{He}}\right)^{\gamma_{inj}-1}\frac{A_{He}}{A_{i}}\; \frac{\alpha\; \bar{X}_{i}^{model,RSG}+\left(1-\alpha\right) \bar{X}_{i}^{model,WR}}{\alpha\; \bar{X}_{He}^{model,RSG}+\left(1-\alpha\right) \bar{X}_{He}^{model,WR}}\;,
$$
\noindent
where $\displaystyle \left(\frac{\bar{X}_{i}}{\bar{X}_{i,He}}\right)^{observed}$ must be equal with the observed flux at
1 TeV:

$$
\left(\frac{\bar{X}_{i}}{\bar{X}_{i,He}}\right)^{observed}=\left(\frac{\phi_{0,i}}{\phi_{0,He}}\right)^{observed}.
$$

In the above equations $\gamma_{inj}$ is the spectral index at injection into the SN shock.

After computations it was clear that, for any value of the $\alpha$
parameter, the data obtained for {\it min}, {\it max} and {\it model/observed} were not in
the limits of the physics hidden behind these factors ({\it min} $\geq 0.5$ and $1.6 \leq$ {\it max} $\leq 2$).

None of our further tries, with similar forms of (\ref{prop}), gave good results (even using $\gamma_{diffusion}=7/3$ or the Table \ref{fit} $\gamma$ values instead of 
$\gamma_{inj}=8/3$).

The formula that worked was:

$$
\left(\frac{\bar{X}_{i}}{\bar{X}_{He}}\right)^{observed}\simeq \left(\frac{Z_{i}}{Z_{He}}\right)^{s}\left(\frac{\bar{X}_{i}}{\bar{X}_{He}}\right)^{model},
$$
\noindent
where {\it s} is one parameter that was determined as being $\sim 1$ (see Table \ref{doi}). The factor $\displaystyle \left(\frac{Z_{i}}{Z_{He}}\right)^{s}$ seems to be related to a different preacceleration 
(and, consequently, a phase space separation) of the wind particles {\bf before their injection into
the supernova shock} of the WR predecessor.

In (\ref{alfa}), trying $\alpha=2/3$ (a contribution for RSG of 66\% and for WR of 33\%),
we obtained for even-Z elements the Table \ref{doi} values.

\begin{table*}
\begin{center}
\begin{footnotesize}
\caption{Mass fraction ratios for even-Z elements when $\alpha=\frac{2}{3}$ and the
preacceleration correction factor is $\left( \frac{Z_{i}}{Z_{He}}\right) ^{s}$.}
\begin{tabular}{c c c c c c c} \hline\hline
elem. & Z & $\left(\frac{\bar{X}_{i}}{\bar{X}_{He}}\right)^{model}$ & $\left(\frac{\bar{X}_{i}}{\bar{X}_{He}}\right)^{observed}$ & {\it s} & factor & $\chi^{2}$ \\
 & & & & & observed/model & \\ \hline
H & 1 & 1.4915 & $0.105\pm 0.075$ & - & $0.9936\div 1.1161$ & - \\
C & 6 & 0.03278 & $0.035\pm 0.015$ & $0.876\pm 0.022$ & 1 & $1\pm 0.1$ \\
O & 8 & 0.0293 & $0.045\pm 0.025$ & $0.998\pm 0.0265$ & 1 & $1.75\pm 0.55$ \\
Ne & 10 & $4.8955\cdot 10^{-3}$ & $(2.22\pm 1.57)\cdot 10^{-2}$ & $1.145\pm 0.022$ & 1 & $0.75\pm 0.02$ \\
Mg & 12 & $1.5336\cdot 10^{-3}$ & $\geq 5.8\cdot 10^{-3}$ & 1 & $6.945\pm 2.936$ & - \\
Si & 14 & $1.9436\cdot 10^{-3}$ & $(5\pm 2)\cdot 10^{-3}$ & 1 & $4.6828\pm 2.02$ & - \\
Fe & 26 & $3.8032\cdot 10^{-3}$ & $\geq 3.05\cdot 10^{-3}$ & 1 & $3.1156\pm 1.7623$ & - \\ \hline
\end{tabular}
\label{doi}
\end{footnotesize}
\end{center}
\end{table*}

\begin{figure*}
\begin{center}
\leavevmode
\includegraphics[width=15cm]{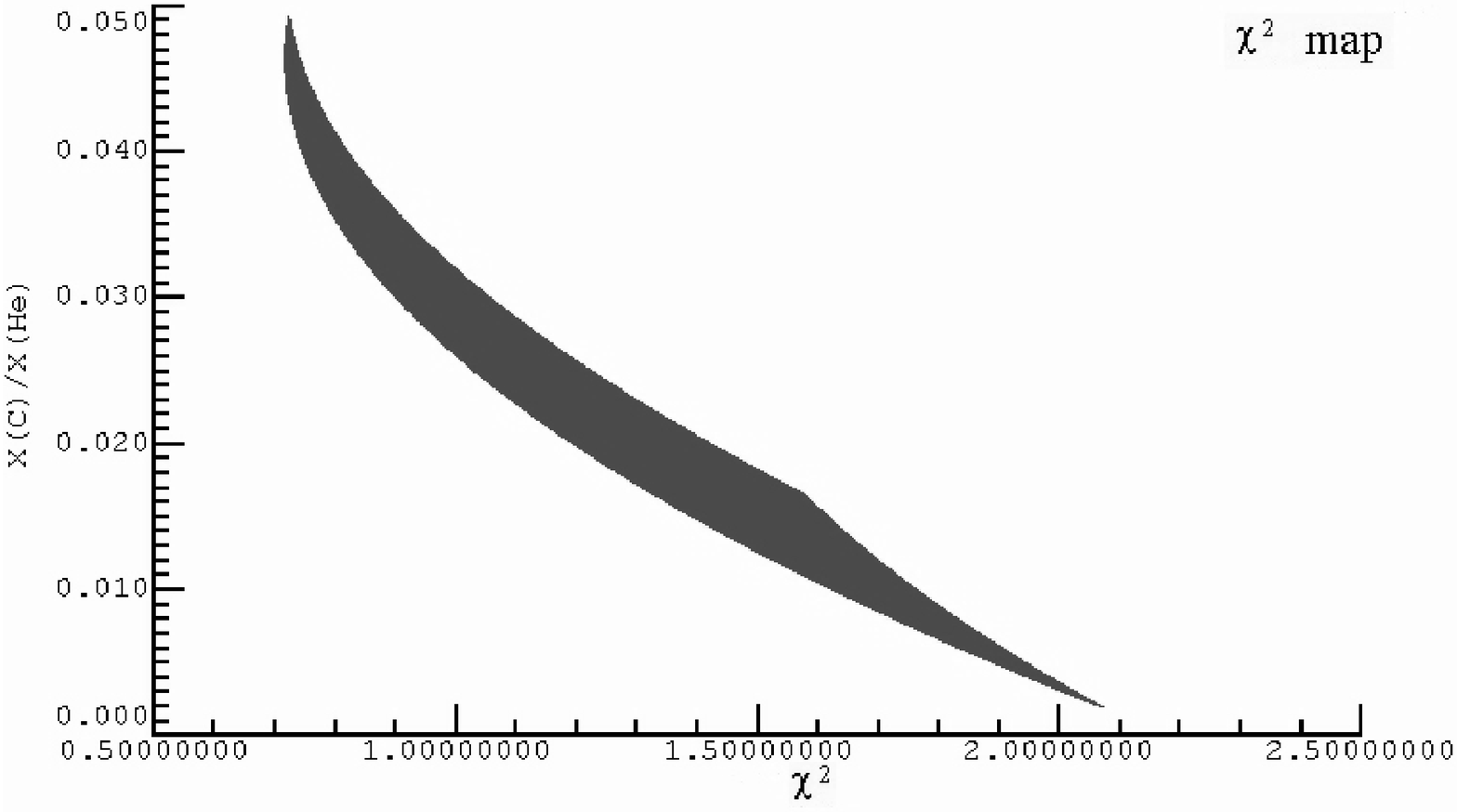}
\caption{$\chi^{2}$ representation of $\left(\frac{\bar{X}_{C}}{\bar{X}_{He}}\right)$ mass fraction ratio for $\alpha=2/3$. In the 
case of $\left(\frac{\bar{X}_{C}}{\bar{X}_{He}}\right)^{model}\simeq \left(\frac{\bar{X}_{C}}{\bar{X}_{He}}\right)^{observed}$, 
$\chi^{2}\simeq 1\pm 0.1$}
\label{chi_c}
\end{center}
\end{figure*}

\begin{figure*}
\begin{center}
\leavevmode
\includegraphics[width=15cm]{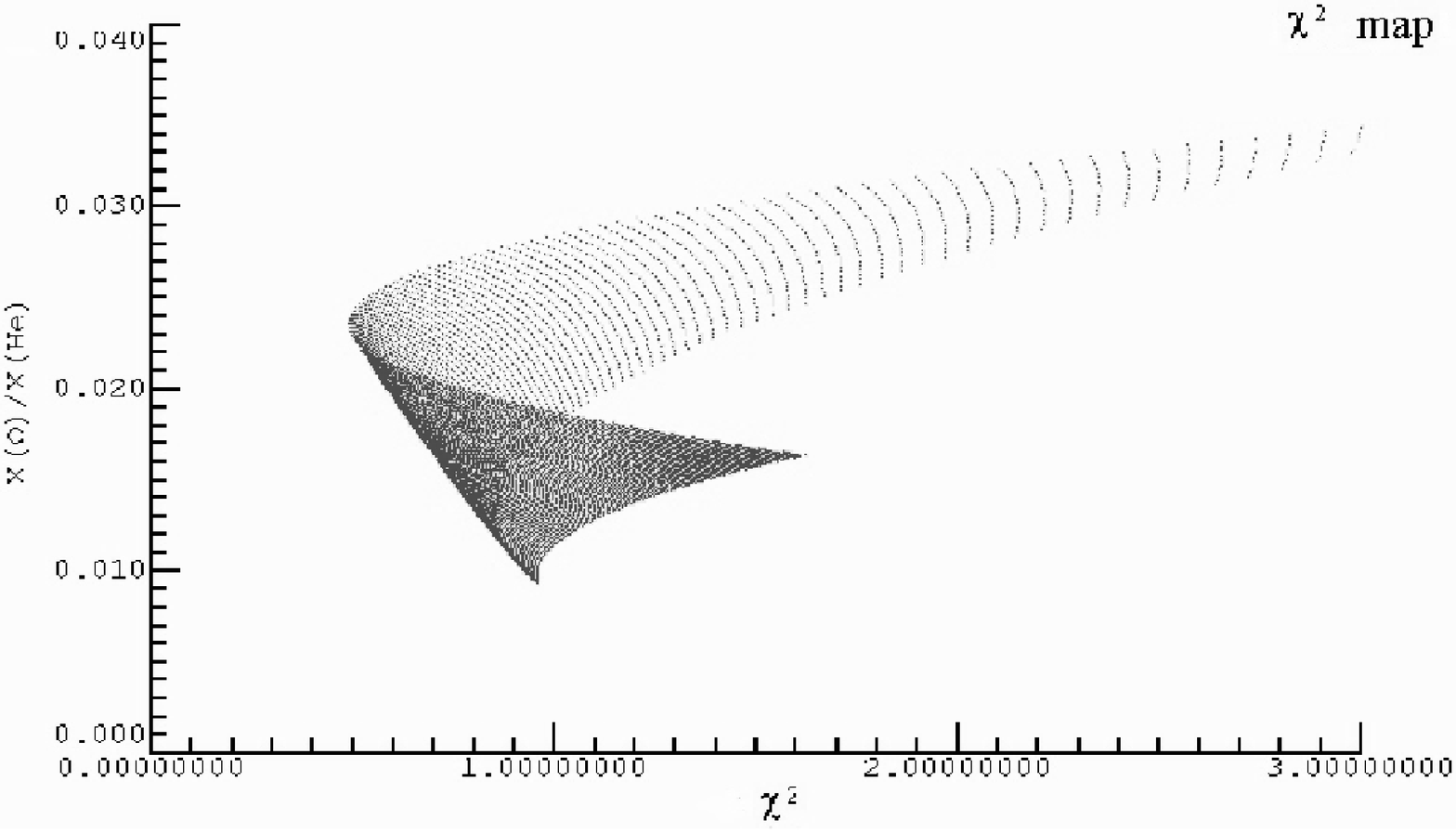}
\caption{$\chi^{2}$ representation of $\left(\frac{\bar{X}_{O}}{\bar{X}_{He}}\right)$ mass fraction ratio for $\alpha=2/3$. In the 
case of $\left(\frac{\bar{X}_{O}}{\bar{X}_{He}}\right)^{model}\simeq \left(\frac{\bar{X}_{O}}{\bar{X}_{He}}\right)^{observed}$, 
$\chi^{2}\simeq 1.75\pm 0.55 $}
\label{chi_o}
\end{center}
\end{figure*}

\begin{figure*}
\begin{center}
\leavevmode
\includegraphics[width=15cm]{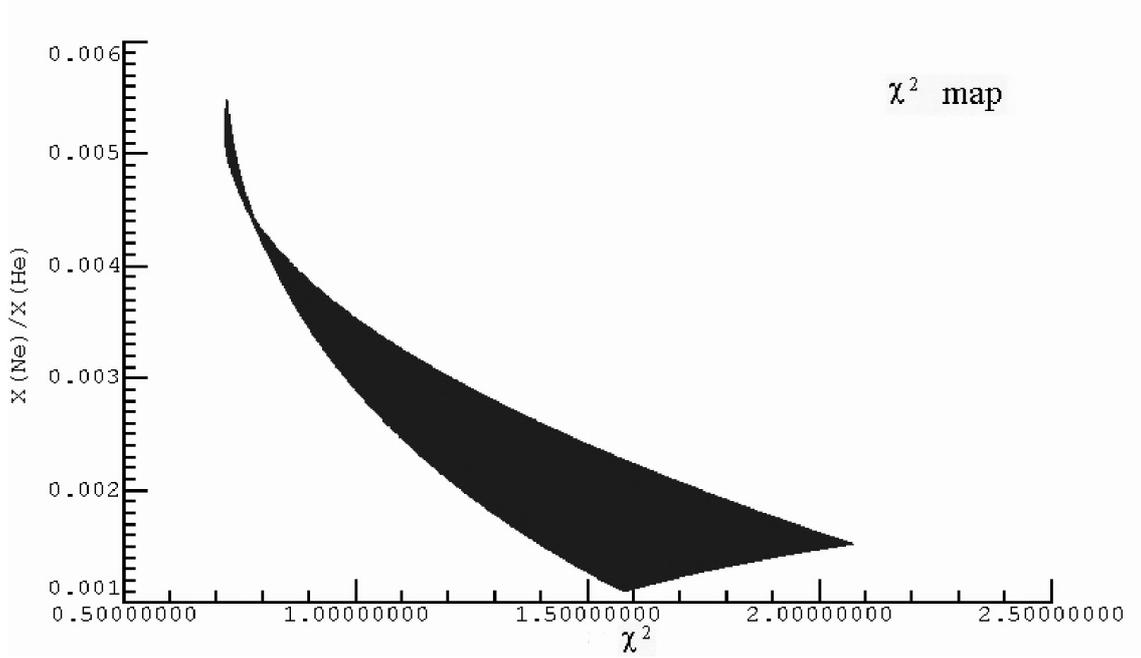}
\caption{$\chi^{2}$ representation of $\left(\frac{\bar{X}_{Ne}}{\bar{X}_{He}}\right)$ mass fraction ratio for $\alpha=2/3$. In the 
case of $\left(\frac{\bar{X}_{Ne}}{\bar{X}_{He}}\right)^{model}\simeq \left(\frac{\bar{X}_{Ne}}{\bar{X}_{He}}\right)^{observed}$, 
$\chi^{2}\simeq 0.75\pm 0.02 $}
\label{chi_ne}
\end{center}
\end{figure*}

Using other $\alpha$ factors we obtained also good results for $\alpha=3/4$, $\alpha=4/5$ and $\alpha=3/5$ (see Tables \ref{trei}, 
\ref{patru} and \ref{cinci}).

\begin{table} [htp]
\begin{center}
\caption{Mass fraction ratios for even-Z elements when $\alpha=\frac{3}{4}$ and the
preacceleration correction factor is $\left( \frac{Z_{i}}{Z_{He}}\right) ^{s}$.  The columns that are missing from this table are
similar with the ones from the $\alpha=\frac{2}{3}$ table.}
\begin{tabular}{c c c c c} \hline\hline
elem. & Z & $\left(\frac{\bar{X}_{i}}{\bar{X}_{He}}\right)^{model}$ & {\it s} & factor \\
& & & & observed/model \\ \hline
H & 1 & 1.52547 & - & $0.9715\div 1.0912$ \\
C & 6 & 0.02439 & $1.041\pm 0.022$ & 1 \\
O & 8 & 0.02785 & $1.022\pm 0.0265$ & 1 \\
Ne & 10 & $4.9329\cdot 10^{-3}$ & $1.142\pm 0.022$ & 1 \\
Mg & 12 & $1.5505\cdot 10^{-3}$ & 1 & $5.4399\pm 1.4418$ \\
Si & 14 & $1.9632\cdot 10^{-3}$ & 1 & $3.6154\pm 0.9561$ \\
Fe & 26 & $3.8316\cdot 10^{-3}$ & 1 & $2.2071\pm 0.8492$ \\ \hline
\end{tabular}
\label{trei}
\end{center}
\end{table}

\begin{table} [htp]
\begin{center}
\caption{Mass fraction ratios for even-Z elements when $\alpha=\frac{4}{5}$ and the
preacceleration correction factor is $\left( \frac{Z_{i}}{Z_{He}}\right) ^{s}$. The columns that are missing from this table are
similar with the ones from the $\alpha=\frac{2}{3}$ table.}
\begin{tabular}{c c c c c} \hline\hline
elem. & Z & $\left(\frac{\bar{X}_{i}}{\bar{X}_{He}}\right)^{model}$ & {\it s} & factor \\
& & & & observed/model \\ \hline
H & 1 & 1.55016 & - & $0.956\div 1.0738$ \\
C & 6 & 0.02015 & $1.148\pm 0.022$ & 1 \\
O & 8 & 0.02722 & $1.033\pm 0.0265$ & 1 \\
Ne & 10 & $4.9751\cdot 10^{-3}$ & $1.138\pm 0.022$ & 1 \\
Mg & 12 & $1.5663\cdot 10^{-3}$ & 1 & $5.2662\pm 1.3628$ \\
Si & 14 & $1.9824\cdot 10^{-3}$ & 1 & $3.4966\pm 0.9014$ \\
Fe & 26 & $3.864\cdot 10^{-3}$ & 1 & $2.1234\pm 0.801$ \\ \hline
\end{tabular}
\label{patru}
\end{center}
\end{table}

\begin{table} [htp]
\begin{center}
\caption{Mass fraction ratios for even-Z elements when $\alpha=\frac{3}{5}$ and the
preacceleration correction factor is $\left( \frac{Z_{i}}{Z_{He}}\right) ^{s}$. The columns that are missing from this table are
similar with the ones from the $\alpha=\frac{2}{3}$ table.}
\begin{tabular}{c c c c c} \hline\hline
elem. & Z & $\left(\frac{\bar{X}_{i}}{\bar{X}_{He}}\right)^{model}$ & {\it s} & factor \\
& & & & observed/model \\ \hline
H & 1 & 1.43411 & - & $1.0334\div 1.1607$ \\
C & 6 & 0.04011 & $0.764\pm 0.022$ & 1 \\
O & 8 & 0.03017 & $0.984\pm 0.0265$ & 1 \\
Ne & 10 & $4.7768\cdot 10^{-3}$ & $1.156\pm 0.022$ & 1 \\
Mg & 12 & $1.4917\cdot 10^{-3}$ & 1 & $8.7264\pm 4.7126$ \\
Si & 14 & $1.892\cdot 10^{-3}$ & 1 & $5.9638\pm 3.3042$ \\
Fe & 26 & $3.7116\cdot 10^{-3}$ & 1 & $4.2764\pm 2.937$ \\ \hline
\end{tabular}
\label{cinci}
\end{center}
\end{table}

We checked also for 1/3, 1/4 and 1 values, but the differences
between the {\it s} values for the triplet C, O and Ne were to big to consider any of these as a
valid candidate for the exponent of Z.

Comparing the ratios {\it observed/model} from the Tables \ref{doi}, \ref{trei}, \ref{patru} 
and \ref{cinci} it can be remarked that, for Mg, Si and Fe the values which enter in the domains of all
three are for $\alpha$ = 2/3 ({\it observed/model} = $4.0088\div 4.878$) and for $\alpha$ = 3/5
({\it observed/model} = $4.01385\div 7.2134$).  Taking C, O and Ne mass fractions as reference ratios for 
{\it CR/surface} and equal to 1, we can see that the behavior of this
ratio versus FIP is having a very good resemblance with the ratio {\it CR/GA} described in
``The FIP Factor Correction'' section of this report. We should remember that it was theoretically predicted a same FIP factor for the elements Mg, Si and Fe. One confirmation of our $\alpha$ choices comes from the fact that it is
known from other theoretical and observational sources that the hydrogen mass fraction is
3 - 4 times less than the total CRs hydrogen mass fraction (in this energy range). This is the reason why, in 
the third column of Tables \ref{doi} - \ref{cinci}, the ratio 
$\displaystyle \left(\frac{\bar{X}_{H}}{\bar{X}_{He}}\right)^{model}$ is so high. The FIP for
hydrogen is 13.6 eV, making from it an element affected by ionization loss and placing it on
the same underabundance (compared with the elements with FIP $\leq 10$ eV) level with C, O
and Ne. But, in both  $\alpha$ = 2/3 and $\alpha$ = 3/5 the {\it observed/model} factor for H is fitting the theory by
 being at least 3 or 4 times smaller than the one for Mg, Si and Fe.

Each one of our two $\alpha$ cases is having its good and bad sides. For $\alpha$ = 2/3 we
have for C, O and Ne the smallest difference in {\it s},
but the sum over all the mass fractions for this $\alpha$ is bigger than the one
for $\alpha$ = 3/5, this sum being closer to 1 for the last case ($\simeq 1.2379$ for
$\alpha$ = 2/3, and $\simeq 1.1065$ for $\alpha$ = 3/5).

The final choice will be made by testing the $^{22}$Ne/$^{20}$Ne and ($^{25}$Mg+$^{26}$Mg)/$^{24}$Mg CRs ratios for both alphas.

\begin{figure*}
\begin{center}
\includegraphics[width=15cm]{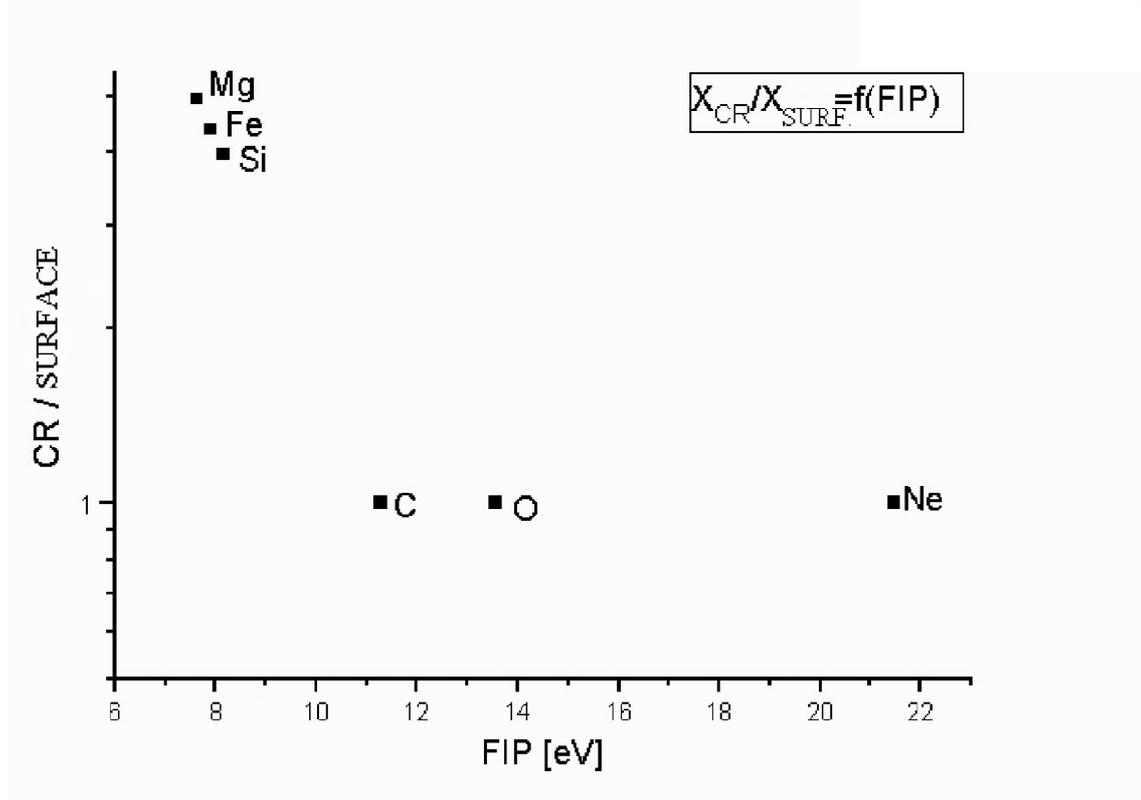}
\leavevmode
\caption{CRs mass fractions to stellar surface mass fractions ratio, corrected for
spallation and transport in the ISM, as function of FIP. Here $\alpha$ is 2/3.}
\label{ptrei}
\end{center}
\end{figure*}

\begin{figure*}
\begin{center}
\includegraphics[width=15cm]{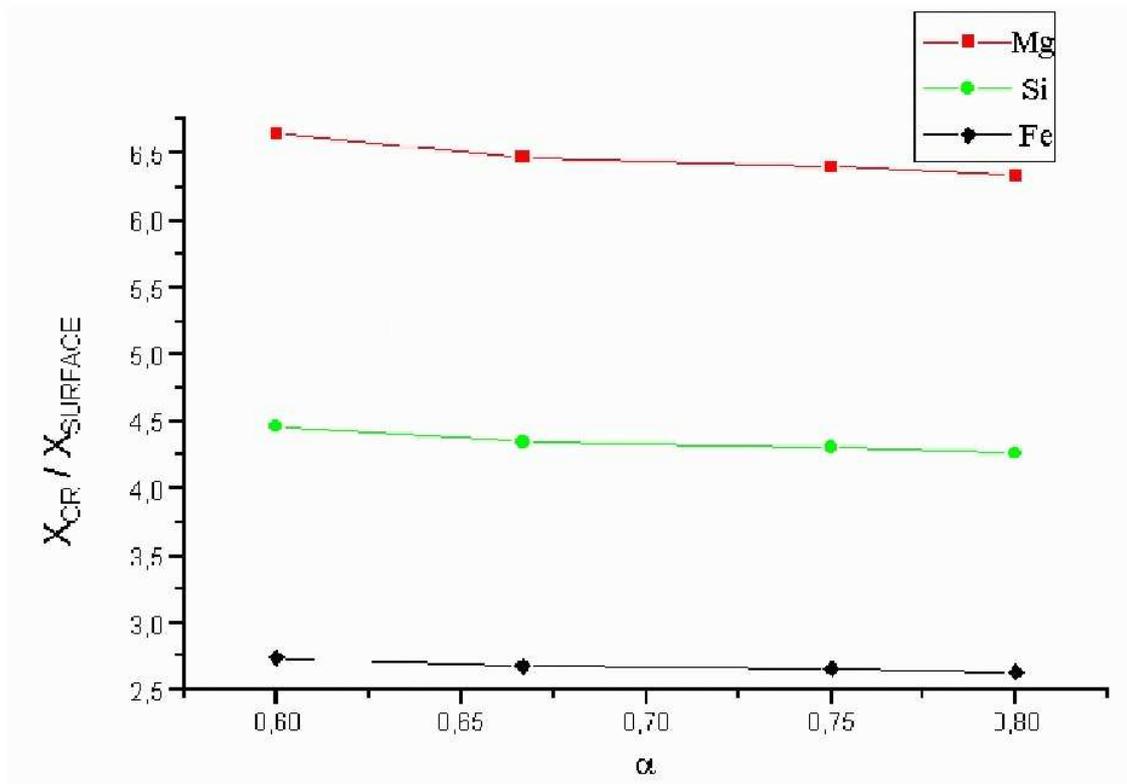}
\leavevmode
\caption{CRs mass fractions to stellar surface mass fractions ratio for Mg, Si and Fe, corrected for
spallation and transport in the ISM, as function of $\alpha$.}
\label{pif}
\end{center}
\end{figure*}

\begin{figure*}
\resizebox{\hsize}{!}{\includegraphics{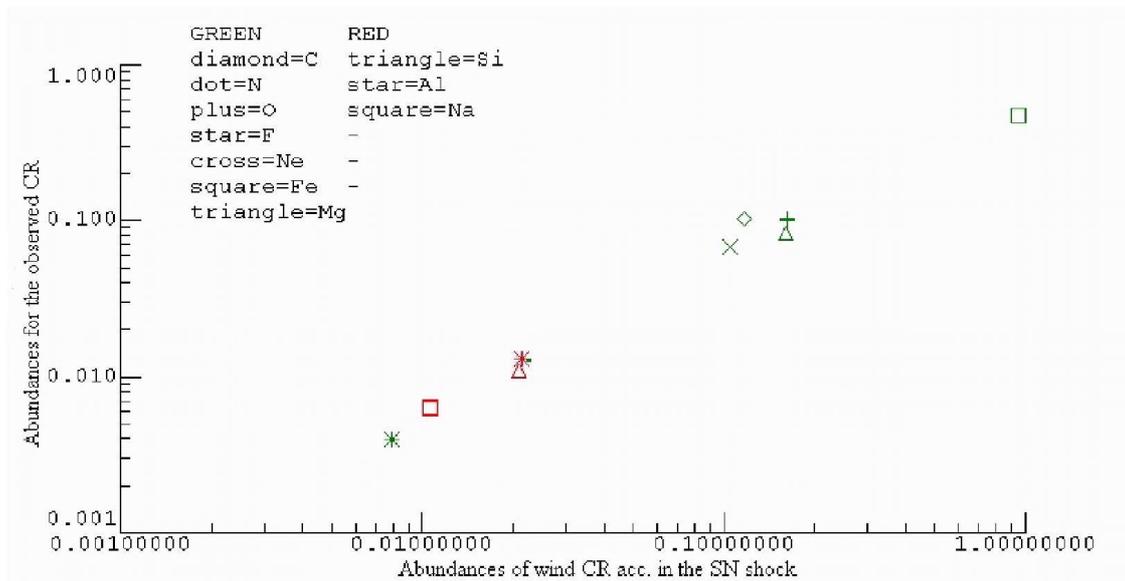}}
\begin{center}
\leavevmode
\caption{The observed ratio between CRs and He mass fraction as function of our model ratio between CRs and He mass fraction.}
\label{etoate}
\end{center}
\end{figure*}

\section{Test}

Neon is one of the most important elements in CRs because its two 
major isotopes, $^{22}$Ne and $^{20}$Ne, are formed by different processes of nucleosynthesis 
($^{23}$Na$(\textnormal{p},\alpha)^{22}$Ne; $^{16}$O$(\alpha,\gamma)^{20}$Ne; $^{14}$N$(\alpha,\gamma)^{18}$F(e$^{+}, \nu)^{18}$O$(\alpha, \gamma)^{22}$Ne). One neon
component measured in carbonaceous chondrites meteorites, called neon-A, is believed to
represent the primordial neon isotopic abundance at the formation of the solar system, and
this value has been adopted as ``standard'' solar system composition: $^{22}$Ne/$^{20}$Ne=0.122
$\pm$ 0.006

The abundance of $^{22}$Ne/$^{20}$Ne measured for the galactic CRs at the orbit of Earth is
0.54$\pm$0.07.
Using standard solar modulation and others cosmic ray propagation models, the CR source ratio
inferred is $^{22}$Ne/$^{20}$Ne=0.38$\pm$0.07 ($^{20}$Ne/$^{22}$Ne=2.6$\pm$0.5) \cite{garcia}.

In our production and propagation of galactic CRs model, the ratio $^{22}$Ne/$^{20}$Ne for
$\alpha$=2/3 is equal to 0.3777, in very good concordance with the above source ratio. 
For  $\alpha$=3/5 this ratio is 0.4772, larger than the CRs source ratio.
We also found that ($^{25}$Mg+$^{26}$Mg)/$^{24}$Mg$=0.3076$ (for $\alpha$=2/3) and
0.3208 (for $\alpha$=3/5) which, compared with the solar system value of 0.27,
suggests that the cosmic ray source and solar system material were synthesized under
different conditions. The obtained values are also suggesting that $\alpha$=2/3 is the right
proportion which we were looking for.

\section{Conclusions}

\begin{itemize}
\item The observed cosmic ray radiation at energy up to $3\times 10^{18}$ eV can be generated by
acceleration of stellar wind particles in supernova shock (see Fig. \ref{etoate}).
\item The contribution ratio of RSG stars and WR stars to the CRs mass fractions is 2:1.
\item Ionization loss is responsable for the underabundance in the observed CR elements with
FIP$\geq 10$ eV.
\item We can consider Si as reference element for overabundance of elements with FIP $\leq 10$ 
eV relative to those affected by ionization loss (see Fig. \ref{ptrei}). The FIP
correction factor will be, in this case, $4.4434\pm 0.4346$. In this way, we can see that the
elements with FIP less than 10 eV are having an larger mass fraction than the elements with FIP
greater than 10 eV (relative to Si), by a factor of $\sim 4$. This happens to be exactly 
$Z_{injection}^2$, the initial degree of ionization squared. Therefore, cosmic ray particles of an
element with an initial degree of ionization of two are more likely to be injected
into the supernova shock.
\item The observed cosmic rays have an increased mass fraction by another
factor of $\displaystyle \left( \frac{Z_{i}}{Z_{He}}\right) ^{s}$. This factor seems to be the result of a differentiated 
radiative preacceleration (phase space dispersion), before shock injection, of different elements found in 
the Wolf-Rayet stellar atmospheres. The reasons which are taking us to this conclusion will be detailed in 
another paper.
\end{itemize}

\ack
For this study we used the programming tool {\bf O-Matrix 5.2}, a very powerfull and fast
programming language. Our complete code is having parts made under {\bf Borland Delphi 5}.

\noindent
Manuscript registered at the Astronomical Institute, Romanian Academy, record No. 248 from 04/19/07.

\end{document}